\documentstyle [twocolumn,epsf]{mn}
\newcommand{\mincir}{\raise
-3.truept\hbox{\rlap{\hbox{$\sim$}}\raise4.truept\hbox{$<$}\ }}
\newcommand{\magcir}{\raise
-3.truept\hbox{\rlap{\hbox{$\sim$}}\raise4.truept\hbox{$>$}\ }}
\newcommand{\minmag}{\raise
-3.truept\hbox{\rlap{\hbox{$<$}}\raise5.truept\hbox{$<$}\ }}
\newcommand{\be}{\begin{equation}}
\newcommand{\ee}{\end{equation}}
\newcommand{\ba}{\begin{eqnarray}}
\newcommand{\ea}{\end{eqnarray}}
\newcommand{\brr}{\begin{array}}
\newcommand{\err}{\end{array}}
\newcommand{\bc}{\begin{center}}
\newcommand{\ec}{\end{center}}

\newcommand{\hm}{\,h^{-1}{\rm Mpc}}

\title[Supercluster properties]
{Supercluster Properties as a Cosmological Probe}
\author[Kolokotronis et al.]{V. Kolokotronis$^{1}$, S. 
Basilakos$^{2}$, M. Plionis$^{1}$. \\
\vspace{0.1cm}
$^1$ Institute of Astronomy \& Astrophysics, National Observatory of Athens, 
I. Metaxa \& V. Pavlou, Palaia Penteli, 15236 Athens, Greece \\
$^2$ Astrophysics Group, Imperial College London, Blackett Laboratory, 
Prince Consort Road, London SW7 2BW, UK\\
}

\begin{document}

\maketitle

\begin{abstract}
We investigate the supercluster shape properties of the all-sky observed
(Abell/ACO) and simulated (Virgo data) cluster catalogues using an approach
based on differential geometry. We reliably identify rich superclusters by
applying percolation statistics to both observed and mock cluster
populations, the latter being constructed following the observational
requirements of the Abell/ACO sample extended out to $z_{\rm max}\leq 0.114$.
We apply a set of shape diagnostics in order to study in a thorough way the
morphological features of superclusters with $\geq 8$ cluster members.
Results demonstrate that filamentarity is the dominant shape feature of
superclusters. On comparing data and models, we show that the $\Lambda$CDM 
($\Omega_{\Lambda}=1-\Omega_{\rm m}=0.7$) model performs better than 
$\tau$CDM, which is excluded at a high confidence level, in agreement with 
other recent large scale structure studies.

\end{abstract}

\vspace{1.0cm}

\section{Introduction}
Superclusters of galaxies possess an eminent position in the structure
hierarchy being the largest mass units we observe today. 
Since they have been seeded by density
perturbations of the largest scale ($\sim 100\hm$), they therefore constitute
objects with which one can study the details of the fluctuations that
gave rise to them (West 1989; Einasto et al. 1997). Additionally, we can
obtain fruitful clues regarding the formation and evolution of the universe 
on the grandest scales and test cosmological models (Bahcall \& Soneira 1984; 
Bahcall 1988; Frisch et al. 1995).

Numerous studies have been devoted not only to delineate the geometrical
pattern of the universe as a whole (Zel'dovich, Einasto \& Shandarin 1982;
de Lapparent, Geller \& Huchra 1991) confirming the picture of a
well designed web-like network, but also to extract statistically complete
supercluster catalogues in order to analyse their spatial distribution and
morphological characteristics (Einasto et al. 2001a, b and references
therein). However, only very recently (Sathyaprakash et al. 1998a; Basilakos,
Plionis \& Rowan-Robinson 2001 hereafter BPR) was any significance given to
cosmological inferences from supercluster shape statistics. Sathyaprakash et
al. (1998a) and BPR have used infrared galaxy samples (1.2 Jy and PSCz
respectively), whereas we presently consider optical clusters (Abell + ACO).

A substantial number of geometrical and topological techniques have been
applied to observational data to explore and assess clustering and
superclustering at large ($> 100\hm$) distances (Mecke et al. 1994;
Sahni \& Coles 1995; Yess \& Shandarin 1996; Dav$\acute{e}$ et al. 1997; 
Kerscher et al. 1997; BPR; Kerscher et al. 2001a, b and references therein).
Nevertheless, an innovative method was devised recently (Sahni 1998; Sahni
et al. 1998; Sathyaprakash et al. 1998b) and has been used with success, so
far, in order to describe in detail the global geometrically complex
features of large scale structure. The first time this new scheme was applied
to astronomical data was in BPR, where it ascertained that the prominent
feature of the large scale structures we see today is filamentarity, as it
has also been observed in N-body simulations of gravitational clustering
(Sathyaprakash et al. 1998b and references therein).

The aim of the present work is not to produce a reliable all-sky supercluster
list. Such attempts have been copiously presented by other authors the last
twenty five years or so, based either on optical or X--ray (ROSAT) cluster
data (Bahcall \& Soneira 1984; Batutski \& Burns 1985; Postman, Huchra \&
Geller 1992; Cappi \& Maurogordato 1992; Plionis, Valdarnini \& Jing 1992;
Zucca et al. 1993; Kalinkov \& Kuneva 1995; Einasto et al. 2001a, b and
references therein). The present analysis will focus on:

\begin{enumerate}

\item whether or not superclusters verify the dominance of filamentarity as 
the basic trait of large scale structure and

\item if so, whether or not supercluster shape and size statistics can be
used to test cosmological scenarios.

\end{enumerate}

The layout of the paper includes the following sections. We first present the 
observed and simulated datasets. In section 3, we discuss 
useful selection parameters of the observed sample, give a brief but 
sufficient report on the method used to investigate supercluster shape 
properties and comment on its stability. In section 4, we derive the
morphological parameters of the real (Abell/ACO) and mock (Virgo)
supercluster data and in section 5 we statistically compare data and dark 
matter (DM) models in an attempt to discriminate between possible structure 
formation scenarios. Finally, our concluding remarks are deserved in section 
6.

\section{The datasets}
\subsection{The Abell/ACO cluster catalogue}
In the present analysis we make use of the combined optical Abell/ACO cluster
list (Abell 1958; Abell, Corwin \& Olowin 1989), which contains 2712 and 1364
entries in the North and South hemispheres respectively but excludes the 1174
supplementary poor southern systems. We are utilising a redshift updated 
version of the catalogue up to a maximum redshift of $z_{\rm max}\le 0.114$ 
and restrict ourselves to $|b|\ge 30^{\circ}$ to avoid severe selection 
biases (cf. sections 2 and 5 of Einasto et al. 2001b) and light absorption by 
the zone of avoidance (ZoA). After taking into account the double cluster 
entries, our final list includes 926 objects (523 Abell and 403 ACO), 
$\sim 80\%$ of which (733/926) have measured redshifts and are also in 
accordance with the above observational limits. Estimated redshifts are taken 
from Plionis \& Valdarnini (1991).

All the above redshifts are heliocentric and transformed to the Local Group
frame using the latest estimates for the Sun's galactic coordinates. 
Redshifts are converted to proper distances using a spatially flat background 
cosmology with $H_{\circ}=100h\,$km$\,$s$^{-1}\,$Mpc$^{-1}$ and 
$q_{\rm \circ}=0.5$. Thus, the redshift cutoff corresponds to a limiting 
distance $R_{\rm max}\le 315h^{-1}\,$Mpc.

\subsection{The Hubble Volume simulations}
For the purpose of our study it is necessary to use the largest up to date 
cosmological N-body simulations. This is the reason for our utilising the 
Virgo Consortium structure formation models (Frenk et al. 2000). Details of 
the simulations have been presented in Colberg et al. (2000) and Evrard et 
al. (2001 and references therein), so we only briefly discuss the main points 
here.

The Hubble Volume simulations follow the evolution of $10^{9}$ particles in
volumes comparable to the whole observable universe. We consider two spatially
flat structure formation models, namely $\Lambda$CDM with
$\Omega_{\Lambda}=0.7$ and $\tau$CDM (massive $\tau$-neutrino contribution).
Further details of the two models are given in Table 1. The lengths of the
boxes are 3$h^{-1}\,$Gpc and 2$h^{-1}\,$Gpc for the two cosmologies
respectively. In both cases, the particle mass is
$m_{\rm p}\ge 2.2\times 10^{12}\,h^{-1}\,M_{\odot}$ and both models are
normalised to the abundance of rich clusters at $z=0$, i.e. $\sigma_{8}=
0.55\,\Omega_{\rm m}^{-0.6}$ (Eke, Cole \& Frenk 1996) and to COBE results.

\begin{table}
\caption[]{Parameters of the DM models.}
\tabcolsep 7pt
\begin{tabular}{ccccccc} 
\hline  
Model & $\Omega_{\rm m}$ & $\Omega_{\Lambda}$ & $h$ & $\Gamma$ & $\sigma_{8}$ & $L_{\rm box}$\\ \hline \hline

$\tau$CDM    &  1.0  &  0.0  &  0.5  &  0.21  &  0.6  & 2$h^{-1}$Gpc \\
$\Lambda$CDM &  0.3  &  0.7  &  0.7  &  0.17  &  0.9  & 3$h^{-1}$Gpc \\ 
\hline     

\end{tabular}
\end{table}

Galaxy clusters are specified using a friend-of-friend algorithm with linking
lengths of order 0.2 and 0.164 for $\tau$CDM and $\Lambda$CDM respectively.
The sizes of simulation boxes certify that an appreciable number of 
independent ``observers'' will be identified within these vast volumes and the 
large number of simulated clusters ($> 10^{6}$) will ensure better statistics 
when supercluster shape properties are measured (see sections 4 and 5).

\section{Geometrical and Shape Formalism}
\subsection{Selection parameters}

The supercluster catalogues for both observed and simulated data are 
constructed 
by using a constant size neighbourhood radius, i.e. a percolation radius
(Zel'dovich et al. 1982), a scheme that has been successfully applied to
similar kinds of studies. We place a sphere of a certain size around each
cluster and then find all neighbouring spheres having an overlap region.
In doing so, we join all clusters falling within the spheres which have
common areas and do this for all clusters in the sample. In order to choose 
the optimal percolation radius, we have repeated the procedure by 
successively increasing the size of the sphere. At the end, we identify the 
radius that yields the maximum number of superclusters, which obviously 
occurs before the percolation of the superclusters themselves. The choice of 
this percolation radius (hereafter $p_{\rm r}$) has been based on criteria 
similar to those of Einasto et al. (1994; 2001a, b and references therein).

To this end, we plot in Figure 1 the
dependence of the number of superclusters on $p_{\rm r}$ for a variety of
percolation radii for both data and models. Solid line corresponds to the
Abell/ACO supercluster data, whereas dashed and dotted lines denote
$\Lambda$CDM and $\tau$CDM respectively. Note, that this plot accounts for
superclusters with at least 8 members, i.e. with multiplicity $k\ge 8$. As
the $p_{\rm r}$ increases, the supercluster number increases accordingly
reaching a maximum at radii between 25$h^{-1}\,$Mpc and
32$h^{-1}\,$Mpc. At larger radii, superclusters start connecting with each
other and at even larger radii, we reach the point of superconnectivity 
with only a few ($\le 5$) huge objects pervading the total volume available. 
We, therefore, choose to work with $p_{\rm r}=27h^{-1}\,$Mpc, which is a 
value well away from the superconnectivity limit ($\ge 33h^{-1}\,$Mpc). 
Notice that our value is consistent with other recent analyses (Einasto et 
al. 1994; Einasto et al. 1997).

\begin{figure}
\mbox{\epsfxsize=8.0cm \epsffile{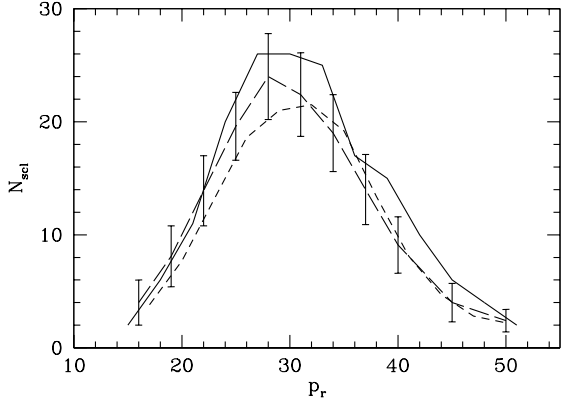}}
\caption{Number of superclusters $N_{\rm scl}$ as a function of $p_{\rm r}$ 
for both data and the two DM models. Solid line denotes the observational 
sample, while dashed and dotted lines correspond to the two models 
($\Lambda$CDM, $\tau$CDM). Note that the superconnectivity limit is similar 
in all three datasets, occurring at depths $>33h^{-1}\,$Mpc. Errorbars 
($\pm 1\sigma$) are shown only for one set of data for clarity.}
\end{figure}

Another observational parameter we wish to compute is the number density of
the Abell and ACO cluster samples as a function of radial depth. This is of 
prime significance since construction of mock cluster catalogues will be 
based on the above parameter. We do so by 
using ten equal volume shells ($\delta V\approx 3\times 10^{6}\,h^{-3}\,$
Mpc$^{3}$) up to $R_{\rm max}\le 315h^{-1}\,$Mpc and calculate the
individual densities of the two observational catalogues separately. The 
corresponding average-over-shells densities for the two samples are 
$n_{\rm Abell}\sim 1.62\times 10^{-5}h^{3}\,$Mpc$^{-3}$ and
$n_{\rm ACO}\sim 2.55\times 10^{-5}h^{3}\,$Mpc$^{-3}$, giving rise to
intercluster separations of order $\sim 39.5h^{-1}\,$Mpc and
$\sim 34h^{-1}\,$Mpc respectively. For completeness, we note that the number
density of the combined catalogue is $\sim 1.82\times 10^{-5}h^{3}\,$Mpc$^{-3}$
corresponding to a mean separation of $\sim 38h^{-1}\,$Mpc in accordance with
other estimates of the same population.

\subsection{The shapefinders}
An appreciable number of geometrical and topological tools have been developed
recently for identifying correctly and quantifying the rich texture of large
scale structure as it appears in large angular or redshift surveys and N-body
simulations (Babul \& Starkman 1992; Luo \& Vishniac 1995; Sahni \& Coles
1995; Sahni 1998; Sathyaprakash et al. 1998b and references therein). In
contrast to traditional schemes (low-order statistics) which bear no relation
to either topology nor shape, the shapefinders have been constructed from
Minkowski functionals (volume $V$, surface area $S$, integrated mean
curvature $C$ and genus $G$) in an attempt to discriminate between complex
morphologies such as filaments (prolate-like objects), pancakes (oblate-like
objects), triaxial structures (ribbons) and spheres (see Figure 1 of Sahni
et al. 1998 and Figure 2 of Sahni 1998 for a schematic presentation of the
above morphological categories).

\begin{figure}
\mbox{\epsfxsize=8.0cm \epsffile{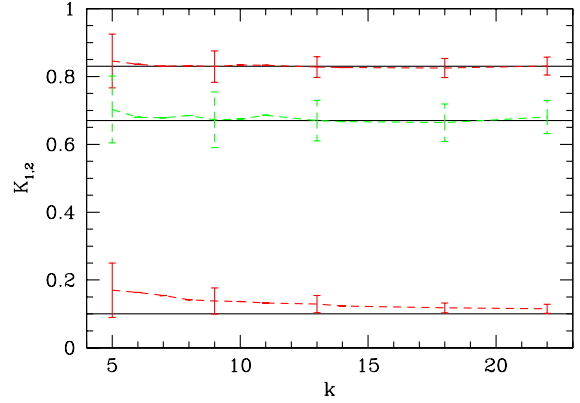}}
\caption{Performance of shapefinders $K_{1}$, $K_{2}$ as a function of
multiplicity $k$. Theoretical values (horizontal solid lines) are compared
with MC computed ones (dashed lines with errorbars) for two ideal objects, 
i.e. a pancake (upper and lower lines denoting $K_{1}$ and $K_{2}$ 
respectively) and a triaxial object (middle line showing $K_{1}$). Errors 
correspond to the $\pm 1\sigma$ scatter from 100 MC realizations performed 
for each $k$, shapefinder and object separately.}
\end{figure}

\begin{figure*}
\mbox{\epsfxsize=13.0cm \epsffile{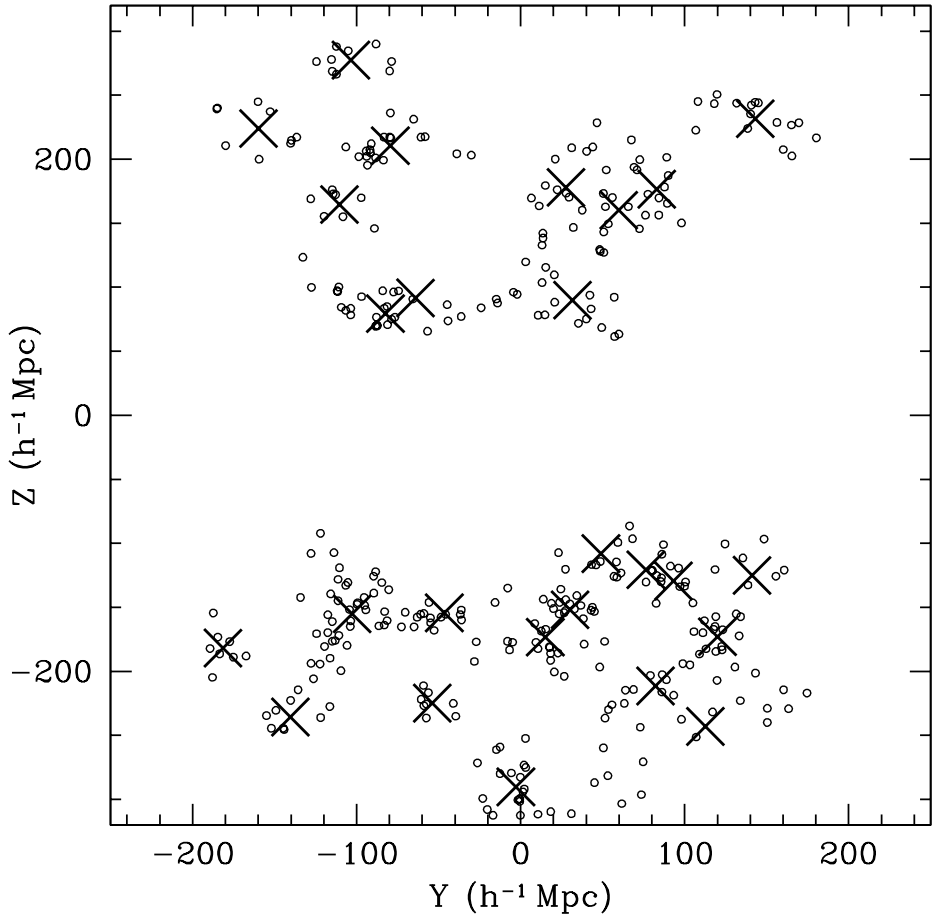}}
\caption{Two dimensional whole sky map of the 26 Abell/ACO superclusters
(crosses). Open symbols denote the clusters associated with superclusters 
with $k\ge 8$. Note, that we have detected 11 superclusters in the North and 
15 in the South.} 
\end{figure*}

For a particular surface of integration class ${\cal C}^{2}$ it is well
known that we can define the four Minkowski functionals which completely
characterise the geometry as well as the topology of a compact surface.
We, thus, introduce the three shapefinders
${\cal H}_{1}=VS^{-1}$, ${\cal H}_{2}=SC^{-1}$ and ${\cal H}_{3}=C$ having 
dimensions of length. Based on these, we can further define the two 
dimensionless shapefinders $K_{1}$ and $K_{2}$ as follows:

\be
K_{1}=\frac{ {\cal H}_{2}-{\cal H}_{1} }{ {\cal H}_{2}+{\cal H}_{1} } 
\ee

and

\be
K_{2}=\frac{ {\cal H}_{3}-{\cal H}_{2} } { {\cal H}_{3}+{\cal H}_{2} } \;\;, 
\ee
where $K_{1,2}\le 1$ by definition. 

Therefore, having identified superclusters 
using the specified percolation radius, shape 
detection is performed only for those objects having $k\geq 8$ members via
the moments of inertia method. By fitting the best triaxial ellipsoid with 
parametric form 

\be
{\mbox{\boldmath$r$}} (\theta,\phi)=I_{1}\,{\rm sin}\theta\,{\rm cos}\phi\,
{\mbox{\boldmath${\hat i}$}}+
I_{2}\,{\rm sin}\theta\,{\rm sin}\phi\,{\mbox{\boldmath${\hat j}$}}+
I_{3}\,{\rm cos}\theta\,{\mbox{\boldmath${\hat k}$}} \;\;, 
\label{eq:parform}
\ee
to real and mock data, we can obtain the three eigenvalues $I_{1}$, $I_{2}$, 
$I_{3}$, which directly relate to the three principle axes of the fitted 
ellipsoid. The volume can be then defined as 
$V=\frac {4\pi}{3}\,I_{1}\,I_{2}\,I_{3}$ and $0\leq \phi\leq 2\pi$, 
$0\leq \theta \leq \pi$. Subsequently, the latter parameters can further be
used to derive the shapefinders ${\cal H}_{\rm i}$ and $K_{\rm i}$.

The above shape technique characterises the shapes of topologically 
non-trivial cosmic objects according to the following classification:

\begin{enumerate}
\item Pancakes if $K_{1}/K_{2}>1$

\item Filaments if $K_{1}/K_{2}<1$

\item Triaxial structures if $K_{1}/K_{2}\approx 1$ and

\item Spheres if $I_{1}=I_{2}=I_{3}$ and $K_{1}\approx K_{2}\approx 0$.
\end{enumerate}

\noindent Ideal filaments (0,1), pancakes (1,0), triaxial structures (1,1)
and spheres (0,0) are represented by the four vertices of the shape plane in
the form of the shape vector $K=(K_{1}$, $K_{2}$), whose amplitude and
direction determines the morphology of an arbitrary 3D surface. Finally, for
shapes close to spherical, $K_{1}$ and $K_{2}$ are very small, so that it is
meaningful to measure their ratio ${\cal R}=K_{1}/K_{2}$ and evaluate
deviations from sphericity (for further details see section 4.2 of BPR).

\subsection{Stability of the method}
The present study is based on superclusters with $k\ge 8$, as quoted above.
This choice has been made on stability grounds and
is a compromise between adequate statistics and efficiency of the shape
technique (see Plionis, Barrow \& Frenk 1991; Jaaniste et al. 1998 their
section 3.2). Indeed, it has been established that superclusters shape
properties are accurately estimated as long as $k\geq 8$ members (Jaaniste
et al. 1998 their figure 2). In order to investigate the performance of our
shape finding method, we utilise a Monte Carlo (MC) approach. In particular,
given as input parameters the three axes $I_{1}$, $I_{2}$ and $I_{3}$ of an
ellipsoid with equation

\be 
\frac{x_{1}^{2}}{I^{2}_{1}}+\frac{x_{2}^{2}}{I^{2}_{2}}+
\frac{x_{3}^{2}}{I^{2}_{3}} \le 1
\ee   
for $k\ge 5$, the question that we answer is ``what are the most probable 
measured $K_{1},K_{2}$ recovered by our shape algorithm?"

In Figure 2, we compare theoretical predictions with MC estimated values of 
$K_{1}$ and $K_{2}$ as a function of multiplicity $k$. Horizontal solid lines
correspond to the input values for an ideal pancake (upper and lower) and an 
ideal triaxial structure (middle) and are taken from Sahni et al. (1998 their 
Table 1). Upper ($K_{1}$) and lower ($K_{2}$) dashed lines with
errorbars denote the derived figures for the pancake, while the middle one 
($K_{1}$) indicates the triaxial object. For the latter case, we observe that 
our MC computed values always approximate extremely well (within $5\%$) the 
input ones for all multiplicities used. Similar results are also obtained 
for $K_{1}$ in the case of the pancake (upper lines). Nevertheless, it is 
evident that our $K_{2}$ estimated figures lie systematically above the input 
ones, albeit asymptotic convergence is gradually restored. This systematic 
trend effectively means that our algorithm tends to make pancakes less 
planar, a feature which is more pronounced at small multiplicities. Exactly the
opposite event occurs in the case of filaments for the input and MC estimated
values of $K_{1}$, thus producing less elongated filaments than it should. 
This effect tends to produce a slightly distorted shape spectrum of 
superclusters, keeping, however, the shape identity of the sample unchanged 
(filaments remain filaments and pancakes remain pancakes). 

We have concluded that, for $k<8$, our input results differ significantly from 
the MC computed values for the study cases of the pancake ($K_{2}$) and the 
filament ($K_{1}$). This is also in line with previous analyses on shape 
determination, thus proving the caveat of the technique for superclusters 
with a few cluster members.

\section{Morphological Properties}
\subsection{The Abell/ACO supercluster features}
We investigate supercluster characteristics according to the observational
requirements and the definitions of shape diagnostics set in section 3.
Taking the optimum value of $p_{\rm r}=27h^{-1}\,$Mpc for the
combined Abell/ACO sample, we end up with 26 superclusters with $k\geq 8$.
The relevant information is presented in Table 2.
In Figure 3, we show a 2D schematic representation of superclusters (crosses) 
superimposed onto the related cluster distribution (open symbols) for the 
observed sample. The absence of objects in the central stripes of the plot is 
due to the ZoA restrictions that we have placed. Note, that only objects with 
$k\geq 8$ are plotted.

\begin{table*}
\caption[]{List of the Abell/ACO superclusters using
$p_{\rm r}=27h^{-1}\,$Mpc. The correspondence of the columns is as follows:
index number, multiplicity, axes of triaxial ellipsoid, distance from us, 
$K_{1}$, $K_{2}$, their ratio ${\cal R}$, the morphological classification 
type and finally the supercluster name (see Einasto et al. 2001a their Table 
A1). F denotes filaments, P is for pancakes and eF and eP symbolise extremely 
elongated or planar superclusters. Note that $I_{1}$, $I_{2}$, $I_{3}$ and 
$R$ have units of $h^{-1}\,$Mpc. We also report that due to our cut in $b$,
only part of the Shapley supercluster has been detected.}

\tabcolsep 9pt
\begin{tabular}{ccccccccccc} 
\hline
Index & $k$ & $I_{1}$ & $I_{2}$ & $I_{3}$ & $R$ & $K_{1}$ & $K_{2}$ &${\cal R}$
& Type & Name\\ \hline \hline 

1  & 16 & 91.6  & 26.8  & 14.1  & 267.9  & 0.1405  & 0.2622  & 0.54 &  F & 
Pegasus-Pisces + Aquarius \\
2  & 22 & 63.0  & 30.6  & 25.1  & 291.0  & 0.0416  & 0.0773  & 0.54 &  F & 
Sculptor \\
3  & 11 & 37.1  & 20.5  & 14.8  & 184.1  & 0.0454  & 0.0644  & 0.71 &  F & 
Pisces - Aries \\
4  & 9  & 31.1  & 23.1  & 12.1  & 129.1  & 0.0723  & 0.0463  & 1.56 &  P &
Pisces \\
5  & 24 & 56.7  & 23.8  & 14.1  & 174.1  & 0.0945  & 0.1429  & 0.66 &  F & 
Pisces - Cetus \\
6  & 8  & 30.8  & 20.4  &  3.1  & 302.4  & 0.4551  & 0.1194  & 3.81 & eP &
Fornax - Eridanus \\
7  & 18 & 84.9  & 19.0  & 11.7  & 117.8  & 0.1206  & 0.3355  & 0.36 & eF & 
Sextans + Leo \\
8  & 8  & 34.5  & 21.2  &  7.7  & 204.3  & 0.1778  & 0.0972  & 1.83 &  P & 
Leo - Sextans \\
9  & 13 & 55.9  & 22.3  & 12.6  & 185.6  & 0.1074  & 0.1614  & 0.67 &  F & 
Ursa Majoris \\ 
10 & 9  & 44.8  & 22.5  & 10.3  & 298.0  & 0.1356  & 0.1218  & 1.11 &  P & 
Leo - Virgo \\
11 & 19 & 63.1  & 34.6  & 17.9  & 225.5  & 0.1004  & 0.0938  & 1.07 &  P & 
Virgo - Coma \\
12 & 9  & 36.1  & 22.1  &  7.7  & 275.3  & 0.1913  & 0.1001  & 1.91 &  P & 
Leo A \\
13 & 16 & 42.0  & 27.1  & 18.2  & 295.5  & 0.0431  & 0.0458  & 0.94 &  F & 
Draco \\
14 & 15 & 47.7  & 33.0  & 13.5  & 135.1  & 0.1333  & 0.0712  & 1.87 &  P & 
Shapley (part) \\
15 & 16 & 55.7  & 37.1  & 20.2  & 184.8  & 0.0742  & 0.0570  & 1.30 &  P & 
Bootes \\
16 & 10 & 35.0  & 14.3  &  9.9  & 210.6  & 0.0713  & 0.1336  & 0.53 &  F & 
Corona Borealis \\
17 & 15 & 56.5  & 19.3  & 13.2  & 116.2  & 0.0835  & 0.1843  & 0.45 & eF & 
Hercules \\
18 & 10 & 37.6  & 13.7  & 10.3  & 240.6  & 0.0680  & 0.1545  & 0.44 & eF & 
Aquarius B \\
19 & 9  & 27.1  & 11.5  &  9.6  & 168.0  & 0.0488  & 0.1048  & 0.47 & eF & 
Aquarius-Cetus \\
20 & 20 & 71.9  & 38.7  & 28.3  & 233.8  & 0.0460  & 0.0681  & 0.68 &  F & 
Aquarius \\
21 & 8  & 33.1  & 18.3  &  5.1  & 190.5  & 0.2757  & 0.1317  & 2.09 & eP & -\\
22 & 46 & 79.6  & 59.3  & 32.4  & 189.7  & 0.0638  & 0.0428  & 1.49 &  P & 
Horologium - Reticulum \\
23 & 8  & 33.9  & 19.4  &  6.6  & 285.2  & 0.2069  & 0.1141  & 1.81 &  P & 
Caelum \\
24 & 12 & 43.5  & 10.9  &  8.0  & 263.3  & 0.0892  & 0.2733  & 0.33 & eF & 
Microscopium \\
25 & 13 & 45.2  & 30.7  & 12.8  & 205.9  & 0.1306  & 0.0726  & 1.80 &  P & -\\
26 & 8  & 31.3  & 17.3  & 10.3  & 277.8  & 0.0737  & 0.0806  & 0.91 &  F & 
Grus \\
\hline

\end{tabular}
\end{table*}

As is evident from Table 2, there are no spherical superclusters in 
concordance with numerical N-body simulations of gravitational clustering and 
similar studies on observed data (Plionis et al. 1992; Sathyaprakash et al. 
1998a, b; Sahni et al. 1998; BPR). More than half of the systems (54$\%$ or
14/26) reveal filamentary structure (cf. section 5.2), 5 of which are dubbed 
as extreme filaments (${\cal R}\leq 0.5$). The rest of the objects (12 or 
46$\%$) are estimated to be pancakes, with 2 of them being extreme cases
(${\cal R}\geq 2$). 

We plot in Figure 4 the results of this study, where the ratio ${\cal R}$ as
a function of $k$ is shown for all 26 real superclusters. The equilibrium 
between filamentarity and planarity, however, seems to be perturbed if we 
extend the $k$ infimum from 8 to 10 members. In this case, we find 71$\%$ 
(12/17) filaments, while for $k\leq 10$ superclusters we obtain 64$\%$ (7/11) 
pancakes. We attribute this to the ill-defined shapes, as evidenced from the 
analysis of section 3.3 (cf. Figure 2). We thus observe that low $k$ 
superclusters are mostly compatible with pancakes or extreme pancakes (see 
object at the upper left of Figure 4), whilst large ones are better described 
as filaments or extreme filaments. Statistically speaking, the larger the 
multiplicity, the more accurately supercluster morphologies are defined, the 
more filaments are revealed.

\begin{figure}
\mbox{\epsfxsize=8.0cm \epsffile{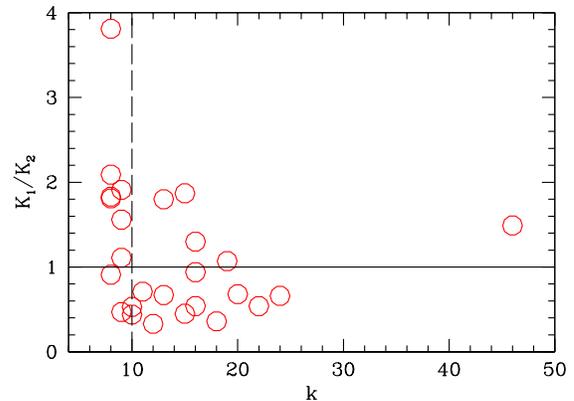}}
\caption{Ratio ${\cal R}$ as a function of multiplicity $k$ for the 26 real 
superclusters. The vertical dashed line corresponds to the $k$-cut discussed 
in text and the horizontal one splits the regions of filaments and pancakes.} 
\end{figure} 

\subsection{The simulated distributions}
The task of constructing mock supercluster catalogues based on the two 
aforementioned DM models ($\tau$CDM, $\Lambda$CDM) is accomplished by using 
the selection parameters of the Abell/ACO population presented earlier. 
Since the lengths of the simulated boxes are $2h^{-1}\,$Gpc and 
$3h^{-1}\,$Gpc for the two models respectively, we can easily define 
several mock cluster distributions as seen by a variety of totally
independent observers. The number of these observers per simulation box can
then be computed as $N_{\rm obs}=(L_{\rm box}/2R_{\rm max})^{3}$.
For $\tau$CDM, we have defined 27 such observers and 64
for $\Lambda$CDM, each of them having exactly the same observational features
as the initial Abell/ACO population (i.e. number density, number of clusters,
selection function, geometry and $R_{\rm max}$).

We, furthermore, make use of the percolation parameter $p_{\rm r}$ that 
maximises the mock supercluster numbers for each model (see Figure 1). For
$\tau$CDM, we have computed $p_{\rm r}=32h^{-1}\,$Mpc and
$p_{\rm r}=28h^{-1}\,$Mpc for $\Lambda$CDM. The latter figures correspond to 
$\sim 21.5\pm 3.5$ superclusters for $\tau$CDM and $\sim 24\pm 4$ for 
$\Lambda$CDM, where means and scatters emanate from the 27 and 64 observers 
respectively.

It is evident that the choice of $p_{\rm r}$ both for the models and the 
observational data is essential not only for the detection of rich 
superclusters, but for getting a better handle on cosmic variance effects as
well (see next section). To this end, we have cross checked our findings 
using the nearest neighbour statistical measure (critical radius $R_{\rm cr}$) 
for clustered distributions (Peebles 2001) 

\be
R_{\rm cr} \simeq \left[\frac{3-\gamma}{\omega_{\rm s} \langle n \rangle 
r_{\circ}^{\gamma}} \right]^{\frac{1}{3-\gamma}}\,\,,
\ee
where $\omega_{\rm s}$ and $\langle n \rangle$ are the solid angle and mean 
number density and $r_{\circ}$, $\gamma$ correspond to the correlation length 
and the slope of the correlation function of the sample under study. Using a 
plausible set of parameters for the clustering properties (correlation 
lengths, amplitudes and slopes) of the two models (see section 3.2 of Colberg 
et al. 2000) and the average values from the literature for the observed 
sample, we find that our figures are very well computed and in excellent 
agreement with those predicted by equation (5).   

\section{Supercluster Statistics}
\subsection{Cosmological implications}
We compare the shape and volume (size) spectrum distributions between the 
data and two cosmological models via a standard Kolmogorov-Smirnov (KS) 
statistical test. We then compute the corresponding probabilities of 
consistency between models and data (${\cal P}_{\rm KS}$) and place these 
results in Table 3. In order to investigate whether the above two statistics 
can be used as a cosmological discriminant, we firstly use the KS test so as 
to compare the two model distributions. It is evident from Table 3 that the 
shape spectrum fails to discriminate between the two models, probably echoing 
the Gaussian initial conditions common to both cosmologies. The volume 
spectrum constitutes a useful tool, since it yields a zero probability 
($\sim 10^{-19}$) that $\tau$CDM and $\Lambda$CDM are being selected  from 
the same parent population.

We display in Figure 5 the results from the shape (left plots) and volume
(right plots) spectrum. Upper panels correspond to the $\tau$CDM model and
lower to $\Lambda$CDM. Hatched areas and open symbols denote the observed
data and filled circles are for the two DM cosmologies. Errorbars in all panels
are the $\pm 1\sigma$ values based on the independent observers and the
vertical solid lines in the left mark the transition limits between
filaments and pancakes. A characteristic worth observing is that in both
models the percentage of filaments exceeds that of pancakes. The latter
finding is in accordance with N-body results (Sathyaprakash et al. 1998a, b). 
A clear indication of the discriminative power of the volume spectrum test
is presented in the right panels of Figure 5. It is apparent that the
$\tau$CDM distribution gives the worst fit to the observed data with
probability of consistency being a mere 0.001. In contrast to this, it seems
that the volume spectrum of $\Lambda$CDM superclusters better represents that
of the Abell/ACO supercluster sample with corresponding probability of
consistency being 0.271. This can be explained by regarding the volume
spectrum as a natural outcome of the different clustering pattern of the two
cosmologies considered here (see Jenkins et al. 1998; Colberg et al. 2000
for a comparison between optical data and Virgo models).

\begin{table}
\caption[]{KS probabilities (${\cal P}_{\rm KS}$) of consistency between
data and models as well as between the two DM models. Note, that in this 
computation, we have used the total number of mock superclusters identified 
within all independent volumes for the two models.}

\tabcolsep 7pt
\begin{tabular}{ccc}
\hline
Pair                     &    Sh. Spectrum     &  Vol. Spectrum \\ \hline 
\hline 
Real - $\Lambda$CDM   &        0.682        &       0.271 \\
Real - $\tau$CDM      &        0.488        &       0.001 \\
$\Lambda$CDM - $\tau$CDM &        0.242        &       $\sim 10^{-19}$\\
\hline
\end{tabular}
\end{table}

To validate our findings, we further explore the selection parameters 
presented in section 3 and test the reliability of our methods. On choosing 
reasonable values for $p_{\rm r}$, the multiplicity infimum and finally a 
smaller set of independent observers, we have recovered that results remain 
remarkably robust. 

\subsection{Optical vs Infrared supercluster shapes}
We attempt to compare the shape spectrum of the optical (Abell/ACO) and the
infrared (PSCz) superclusters in order to probe whether or not the prominence
of filamentarity is retained by all-sky data with diverse global geometrical
and selection properties. For this purpose, we have resorted to the recently
constructed all-sky PSCz supercluster catalogue (BPR) counting $\sim 12000$
infrared galaxies.

We display in Figure 6 the shape distribution of infrared and optical data
(hatched histograms) for superclusters with $k\ge 8$ (left) and $k\ge 10$
(right). It is obvious that the filamentarity dominance is confirmed by both
kinds of data, although it is less prominent in our optical sample for
$k\ge 8$ (cf. Figure 4 and discussion in section 4.1). In fact, for $k\ge 10$,
we observe that infrared and optical data yield almost identical values for
the fractions of filaments ($\sim 70\%$) and pancakes ($\sim 30\%$).

A final word of caution is due here. The above argument is not to cast any
shadow on the results based on the use of the nominal $k$ infimum applied to 
the present analysis. Had we used superclusters with $k\ge 10$, we would have 
obtained qualitatively the same results but run the risk of being too close 
to the validity limit of the KS test.

\begin{figure}
\mbox{\epsfxsize=8.0cm \epsffile{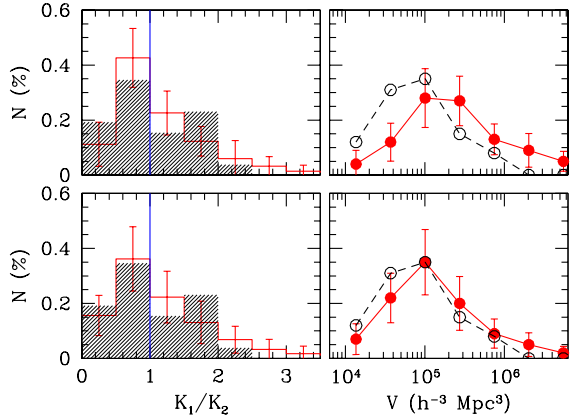}}
\caption{Statistical comparison between data (hatched areas and open
circles) and models (filled symbols). Shape spectra are plotted in left
panels and multiplicity functions in right for $\tau$CDM (top) and
$\Lambda$CDM (bottom) respectively.} 
\end{figure}

\begin{figure}
\mbox{\epsfxsize=8.0cm \epsffile{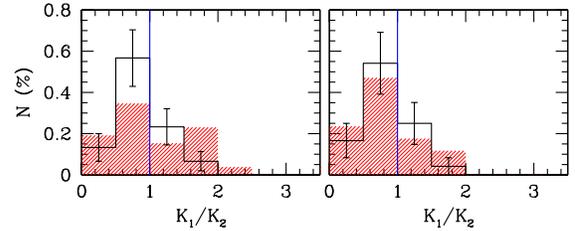}}
\caption{Shape distributions for Abell/ACO (hatched areas) and PSCz
superclusters for $k\geq 8$ (left) and $k\geq 10$ (right). Poissonian errors
are only shown for the PSCz sample for clarity.}
\end{figure}

\section{Concluding Remarks}
We have explored the shape parameters of superclusters identified in observed 
optical (Abell/ACO) and simulated (Virgo DM models) cluster populations, up 
to a depth of $315h^{-1}\,$Mpc. We have applied a constant size percolation 
radius $p_{\rm r}$ to both data and DM models in order to detect 
superclusters within the volumes of the given cluster samples. Using the
selection features of the Abell/ACO catalogue, we have constructed several 
mock supercluster samples based on independent observers for the two 
cosmological models ($\tau$CDM and $\Lambda$CDM). The investigation of 
supercluster morphologies has been based on a differential geometry method 
devised by Sahni et al. (1998a).

Results indicate that filamentary structures dominate the volumes considered 
here over pancake morphologies, as seen in both data and models. This, further
supports the idea that filamentarity is the main structural feature of large
scale structure, since it has been reliably detected in all available
infrared and optical samples analysed up to date. No spherical superclusters
are detected in accordance with similar observational analyses and
theoretical predictions from large N-body experiments (Sathyaprakash et al.
1998a; BPR).

A standard KS statistical test of consistency between data and the two
models on the size distributions has clearly given a preference to the
$\Lambda$CDM ($\Omega_{\Lambda}=0.7$) cosmology. On the contrary, $\tau$CDM
is rejected at a high ($99.9\%$) significance level. Finally, the shape
spectrum is insensitive to the two cosmologies, reflecting their common 
Gaussian initial conditions. The latter findings are in concordance with a 
similar analysis based on infrared PSCz supercluster properties.

\section* {Acknowledgements}
We have made use of two cosmological models extracted from the Virgo
Consortium simulations following the public release of the available data
(Frenk et al. 2000). The Hubble Volume cluster catalogues can be found in 
http://www.mpa-garching.mpg.de/Virgo/.


{\small
\begin{thebibliography}{}
\bibitem[]{}Abell G., 1958, ApJS, 3, 211
\bibitem[]{}Abell G., Corwin H., Olowin R., 1989, ApJS, 70, 1
\bibitem[]{}Babul A., Starkman G., 1992, ApJ, 401, 28
\bibitem[]{}Bahcall N., Soneira R., 1984, ApJ, 277, 27 
\bibitem[]{}Bahcall N., Ann. Rev. Astr. Ap., 26, 631
\bibitem[]{}Basilakos S., Plionis M., Rowan-Robinson M., 2001, MNRAS, 223, 
47 (BPR)
\bibitem[]{}Batutski D., Burns J., 1985, AJ, 90, 1413
\bibitem[]{}Cappi A., Maurogordato S., 1992, A\&A, 259, 423
\bibitem[]{}Colberg J. et al., 2000, MNRAS, 319, 209
\bibitem[]{}Dav$\acute{e}$ R., Hellinger D., Primack J., Nolthenius R., 
Klypin A., 1997, MNRAS, 284, 607
\bibitem[]{}de Lapparent V., Geller M. J., Huchra J. P., 1991, ApJ, 
369, 273 
\bibitem[]{}Einasto M., Einasto J., Tago E., Dalton G., Andernach H., 1994, 
MNRAS, 269, 301
\bibitem[]{}Einasto M., Tago E., Einasto J., Andernach H., 1997, A\&AS, 123, 
119
\bibitem[]{}Einasto M., Einasto J., Tago E., Mueller V., Andernach H., 2001a, 
astro-ph/0012536
\bibitem[]{}Einasto M., Einasto J., Tago E., Andernach H., Dalton G., 
Mueller V., 2001b, astro-ph/0012538
\bibitem[]{}Eke V., Cole S., Frenk C. S., 1996, MNRAS, 282, 263
\bibitem[]{}Evrard A. et al. 2001, ApJ, {\it submitted}, astro-ph/0110246 
\bibitem[]{}Frenk C. et al., 2000, MNRAS, astro-ph/0007362
\bibitem[]{}Frisch P. et al., 1995, A\&A, 296, 611
\bibitem[]{} Jaaniste J., Tago E., Einasto M., Einasto J., Andernach H., 
Mueller  V., 1998, A\&A, 336, 35
\bibitem[]{}Jenkins A. et al., 1998, ApJ, 449, 20
\bibitem[]{}Kalinkov M., Kuneva I., 1995, A\&AS, 113, 451
\bibitem[]{} Kerscher M. et al., 1997, MNRAS, 284, 73
\bibitem[]{}Kerscher M., Mecke K., Schmalzing J., Beisbart C., Buchert T., 
Wagner H., 2001a, A\&A, 373, 1
\bibitem[]{}Kerscher M. et al., 2001b, A\&A, 377, 1
\bibitem[]{}Luo S., Vishniac E., 1995, ApJS, 96, 429
\bibitem[]{}Mecke K., Buchert T., Wagner H., 1994, A\&A, 288, 697
\bibitem[]{}Peebles P. J. E., 2001, ApJ, 557, 495
\bibitem[]{}Plionis M., Valdarnini R., 1991, MNRAS, 249, 46
\bibitem[]{}Plionis M., Barrow J.D., Frenk C.S., 1991, MNRAS, 249, 662
\bibitem[]{}Plionis M., Valdarnini R., Jing Y. P., 1992, ApJ, 398, 12
\bibitem[]{}Postman M., Huchra J., Geller M., 1992, ApJ, 384, 404
\bibitem[]{}Sahni V., 1998, in Sato K., editor, Proceedings of the IAU 
symposium No. 183, Kyoto, Japan, astro-ph/9803189
\bibitem[]{}Sahni V. \& Coles P., 1995, Phys. Rep., 262, 1
\bibitem[]{}Sahni V., Sathyaprakash B. S., Shandarin S., 1998a, ApJ, 495, L5
\bibitem[]{}Sathyaprakash S. B., Sahni V., Shandarin S., 1998b, ApJ,
508, 551
\bibitem[]{}Sathyaprakash S. B., Sahni V., Shandarin S., Fisher B. K., 1998a, 
ApJ, 507, L109 
\bibitem[]{}West J. M., 1989, ApJ, 347, 610
\bibitem[]{}Yess C. \& Shandarin S., 1996, ApJ, 465, 2
\bibitem[]{}Zeldovich, Ya. B., Einasto, J., Shandarin, S., 1982, Nature,
300, 407
\bibitem[]{}Zucca E., Zamorani G., Scaramella R., Vettolani G., 1993, ApJ, 
407, 470
\end{thebibliography}
}
\end{document}